\documentclass[fleqn,usenatbib,useAMS]{mnras}
\usepackage[table,xcdraw]{xcolor}
\usepackage{newtxtext,newtxmath}
\usepackage{graphicx}
\usepackage[utf8]{inputenc}
\usepackage[english]{babel}
\usepackage{subfigure}
\usepackage[T1]{fontenc}
\usepackage[T1]{fontenc}
\usepackage{hyperref}
\usepackage{rotating}
\usepackage{soul}
\usepackage[flushleft]{threeparttable}
\hypersetup{colorlinks,citecolor=blue,filecolor=blue,linkcolor=blue,urlcolor=blue}
\DeclareRobustCommand{\VAN}[3]{#2}
\let\VANthebibliography\thebibliography
\def\thebibliography{\DeclareRobustCommand{\VAN}[3]{##3}\VANthebibliography}

\usepackage{graphicx}	% Including figure files
\usepackage{amsmath}	% Advanced maths commands
\title[Twin Radio galaxies]{The Twin Radio galaxy TRG J104454+354055 }
\author[Gopal-Krishna~et~al.]{
Gopal-Krishna$^{1}$,\thanks{E-mail: gopaltani@gmail.com}
Ravi Joshi$^{2,6}$,\thanks{E-mail: rvjohshirv@gmail.com}
Dusmanta Patra$^{3}$,
Xiaolong Yang$^{4,5}$,
Luis C. Ho$^{6,7}$, 
Paul J.\ Wiita$^{8}$,
\newauthor
Amitesh Omar$^9$
\\
$^{1}$UM-DAE Centre for Excellence in Basic Sciences, Mumbai, \it{400098}; India\\
$^{2}$Indian Institute of Astrophysics, Koramangla, Bangalore,  \it{560034}; India\\
$^{3}$S N Bose National Centre for Basic Sciences, Kolkata, \it{700106}; India\\
$^{4}$Shanghai Astronomical Observatory, Key Laboratory of Radio Astronomy, Chinese Academy of Sciences, Shanghai, \it{200030}, China \\
$^{5}$Shanghai Key Laboratory of Space Navigation and Positioning Techniques, Shanghai Astronomical Observatory, Chinese Academy of Sciences, Shanghai, \it{200030}, China \\
$^6$Kavli Institute for Astronomy and Astrophysics, Peking University, Beijing, \it{100871}, China\\
$^7$Department of Astronomy, School of Physics, Peking University, Beijing, \it{100871}, China \\
$^{8}$Department of Physics, The College of New Jersey, 2000 Pennington Rd., Ewing, NJ, \it{08628-0718}, USA\\
$^{9}$Aryabhatta Research Institute of observational sciencES, Nainital, \it {263002}; India\\
}
\date{Accepted XXX. Received YYY; in original form ZZZ}
\pubyear{2022} 
\begin{document}
\label{firstpage}
\pagerange{\pageref{firstpage}--\pageref{lastpage}}
\maketitle

% Abstract of the paper
\begin{abstract}

We report observations of a bright dumb-bell system of galaxies ($z =0.162$) with the  upgraded Giant Metrewave Radio Telescope ($u$GMRT), which show that each member of this gravitationally bound pair of galaxies hosts  bipolar radio jets extended on  100 kiloparsec scales. Only two cases of such radio morphology have been reported previously, both being dumb-bell systems, as well. The famous first example, 3C 75, was discovered 4 decades ago, and the second case was discovered 3 decades ago. This implies that such `Twin-Radio-Galaxies' (TRGs) are an exceedingly rare phenomenon. As in the case of its two senior cousins, the bi-polar radio jets of the present TRG (J104454+354055) exhibit strong wiggles and are edge-darkened (Fanaroff-Riley class I). However, there are important differences, too. For instance, the jets in the present TRG do not merge and, moreover, show no signs of distortion due to an external crosswind. This makes the present TRG a much neater laboratory for studying the physics of (sideway) 
colliding jets of relativistic plasma. This TRG has a Wide-Angle-Tail (WAT) neighbour hosted by another bright galaxy belonging to the same group, which appears to be moving towards the TRG.
\end{abstract}

% Select between one and six entries from the list of approved keywords.
% Don't make up new ones.
 
%%%%%%%%%%%%%%%%%%%%%%%%%%%%%%%%%%%%%%%%%%%%%%%%%%

%%%%%%%%%%%%%%%%% BODY OF PAPER %%%%%%%%%%%%%%%%%%

\section{Introduction}
\label{sec_intro}
Galaxy mergers are the bedrock of the evolving cosmic structure leading to hierarchical galaxy formation \citep[e.g.,][]{Barnes1996ApJ...471..115B, Bell2006ApJ...652..270B}. 
Since practically all large galaxies are believed to harbour a supermassive black hole (SMBH) \citep{1998Natur.395A..14R, Kormendy2013ARA&A..51..511K}, %{\color{red}Kormendy \& Ho 2013, ARA\&A 51, 511)}, 
the possibility of SMBH coalescense due to galaxy mergers
\citep[e.g.,][and references therein]{Begelman1980Natur.287..307B, Gualandris2017MNRAS.464.2301G} has become a driver for the rapidly advancing new field of gravitational-wave astronomy \citep[e.g.,][]{Hughes2003AnPhy.303..142H, Amaro-Seoane2017arXiv170200786A, Jenkins2022arXiv220205105J}. From an observational viewpoint, if one or both members in the merging galaxy pair host a radio-loud Active Galactic Nucleus (AGN), this can serve as a direct tracer of the SMBH merger process, via monitoring by Very Long Baseline Interferometry (VLBI). The prospects become much more exciting when {\it both} AGN are radio loud \citep[][]{Owen1985ApJ...294L..85O,Simpson2002MNRAS.334..511S}. The main techniques used to search for SMBH binaries (i.e., dual AGN) include photometric and spectroscopic measurements and radio interferometric imaging which is particularly promising because of the high resolution it can provide \citep[e.g.,][]{Komossa2003ApJ...582L..15K,Liu2010ApJ...715L..30L, Bogdanovic2015ASSP...40..103B,Shangguan2016ApJ...823...50S,Rubinur2021MNRAS.500.3908R}. However, the success rate in finding merging SMBHs with discernible radio jets has been minuscule \citep[e.g.][]{Burke-Spolaor2011MNRAS.410.2113B, Tremblay..10.1093/mnras/stw592}, with just a few reasonably compelling cases reported so far.  \par

On the smallest scale (< 0.1 pc), the BL Lac object OJ 287 remains the only strong case for hosting an SMBH binary, as gleaned from its (paired) optical outbursts, recurring every $\sim 12$ years \citep{Sillanpaa1988ApJ...325..628S}. Its long-term optical light-curve has been modelled in terms of a SMBH binary with an orbital semi-major axis of 0.05 pc, whose secondary crosses the accretion disk of the primary SMBH twice during each orbit 
(e.g., \citealt{Lehto1996ApJ...460..207L, Valtonen2016ApJ...819L..37V, Dey2021MNRAS.503.4400D}; see, also \citealt{Britzen2018MNRAS.478.3199B}). Even on slightly larger 
(parsec) scale, only a couple of dual radio-AGN candidates seem genuine, the most prominent being the source 0402+379 for which VLBI has yielded a SMBH binary separation of $\sim$7 pc \citep[][]{Rodriguez2006ApJ...646...49R, Burke-Spolaor2011MNRAS.410.2113B, Bansal2017ApJ...843...14B}. More recently, a dual radio-AGN, with a SMBH separation of just 0.35 pc has been found in the Seyfert galaxy NGC 7674 at $z$ = 0.0289 \citep{Kharb2017NatAs...1..727K}. \par 

It is worth noting that while assessing the genuineness of such radio-loud cases, one needs to guard against the possibility that the candidate dual radio-AGN may turn out to be just (i) a pair of knots in a VLBI jet, or (ii) hot spots in a single compact-symmetric radio source \citep[see, e.g.,][]{Wrobel2014ApJ...792L...8W}, or (iii) even a case of (superluminal) gravitational milli-lensing \citep[]{Gopal1996A&A...315..343G}. Naturally, the binarity becomes much less ambiguous when the two radio nuclei are separated on kilo-parsec scale (permitting detailed radio imaging), even though such SMBH pairs are clearly remote from an eventual merger. So far, just two cases of gravitationally bound pairs of radio galaxies with kiloparsec-scale jets have been identified, each hosted by a pair of similarly bright elliptical galaxies within a common envelope ("dumb-bell"). The famous first example of such "Twin Radio Galaxy (TRG)" is the proto-type 3C 75, a  
 Wide-Angle-Tail (WAT) radio source of overall size $\sim$0.5 Mpc, located near the centre of the galaxy cluster Abell 400 ($z$ = 0.023). It was discovered almost 4 decades ago \citep{Owen1985ApJ...294L..85O}, which testifies to the extreme rarity of this cosmic phenomenon. The second TRG, which was discovered 3 decades ago and has received much less attention, is the source PKS 2149-158 of size $\sim$400 kpc, located near the outskirts of the cluster Abell 2382 ($z$ = 0.062) \citep{Parma1991AJ....102.1960P, Guidetti2008A&A...483..699G}.

In this paper we report the discovery of the third case of a bound pair of galaxies, each of which has a bi-polar radio jet extended $>$ 100 kpc scale. The  dumb-bell galaxy lies at the centre of a galaxy group at  $z \sim 0.16$ \citep{Yuan2016MNRAS.460.3669Y}.  This source was initially classified as a X-shaped radio galaxy  candidate~\citep{2019ApJS..245...17Y, Joshi2019ApJ...887..266J}, but in a follow-up programme with uGMRT, it was found to be a TRG. Here we report radio-optical properties of this TRG and briefly compare it with the other two TRGs mentioned above.
Throughout this paper a cosmology with ${H}_{0}\,=70\,\mathrm{km}\,{{\rm{s}}}^{-1}\,{\mathrm{Mpc}}^{-1}$, ${{\rm{\Omega }}}_{{\rm{m}}}=0.3$ and ${{\rm{\Omega }}}_{{\rm{\Lambda }}}=0.7$ is adopted. This gives a luminosity distance of 780 Mpc for the TRG and a scale of 2.797 kpc per arcsecond at the source redshift of $z=0.16278$.

\section{Observations and analysis}

 The inset in Fig. 1a shows a $g,r,z$ colour composite image from the  Dark Energy Camera Legacy  survey (DECaLS) \citep{Dey2019AJ....157..168D}, of the dumb-bell consisting of the galaxies `A' \& `B', both with $z \simeq 0.162$, and a third, fainter galaxy 'C' of unknown redshift, within the same stellar halo. The colour composite image of another bright galaxy `D', which has a redshift of 0.160 and is located 2.72 arcmin (= 456 kpc) west of the dumb-bell is shown in the inset of Fig. 1c. Particulars of these 4 galaxies are given in Table~\ref{tab:spec}.   Low-resolution optical spectroscopic observations of the dumb-bell system were carried out  using the HFOSC instrument mounted on the 2m-Himalayan Chandra Telescope \citep{Prabhu2014PINSA..80..887P}, with grisms Gr\#7 (3500–7800~\AA) and slit-width of 2.0 arcsec. The data were reduced with the standard IRAF\footnote{IRAF is distributed by the National Optical Astronomy Observatories, which are operated by the Association of Universities for Research in Astronomy, Inc., under cooperative agreement with the National Science Foundation.} package. The optical spectra were modelled using the penalized PiXel-Fitting method  (pPXF, \citealt{Cappellari2004PASP..116..138C, Cappellari2017MNRAS.466..798C}) and the measured stellar velocity dispersions of  $\sigma_* = 412.2\pm 34.8 \rm\  km~s^{-1}$ and $254.5 \pm 20.4 \rm\ km~s^{-1}$ for the galaxies A and B were  used to estimate the black-hole masses \citep[][see their table 5]{Greene2020ARA&A..58..257G}  (see Table~\ref{tab:spec}). The non-detection of [O {\sc iii}]$\rm \lambda 5007\AA$ nebular emission line for both galaxies implies that they are  low-excitation radio galaxies (LERGs) \citep{Best2012MNRAS.421.1569B}.

The radio observations were made using the upgraded Giant Metrewave Radio Telescope ($u$GMRT), in the band-5  (900 to 1450 MHz), with 400 MHz bandwidth encompassing 2048 channels  \citep{Swarup1991CSci...60...95S, Gupta2017CSci..113..707G}. The data were reduced following the standard procedures, using the Astronomical Image Processing System ({\tt AIPS})\footnote{http://www.aips.nrao.edu}. While calibrating the data, bad data were flagged out at various stages. The data for antennas with large errors in antenna-based solutions were checked and flagged over the requisite time ranges. The flux calibration was performed using 3C286 and the flux density scale defined by \citet{Perley2013ApJS..204...19P}. The phase calibration was done using the source 1033+395\footnote{http://http://www.vla.nrao.edu/astro/calib/manual/csource.html}.
The final image was made after several rounds of phase self-calibration, and one round of amplitude self-calibration. An on-source time of 48 minutes resulted in an rms noise of  34 $\mu$Jy/beam, where the beamsize is $3.15^{''} \times 1.73^{''} $ (P.A. = 71 deg) (Fig. \ref{fig.trg_ugmrt}).

\section{Results}
\label{sect.res}

\begin{figure*}
   \includegraphics[width=16.5cm,height=9.2cm,angle=0, trim={0.0cm 0.0cm 0.0cm 0cm},clip]{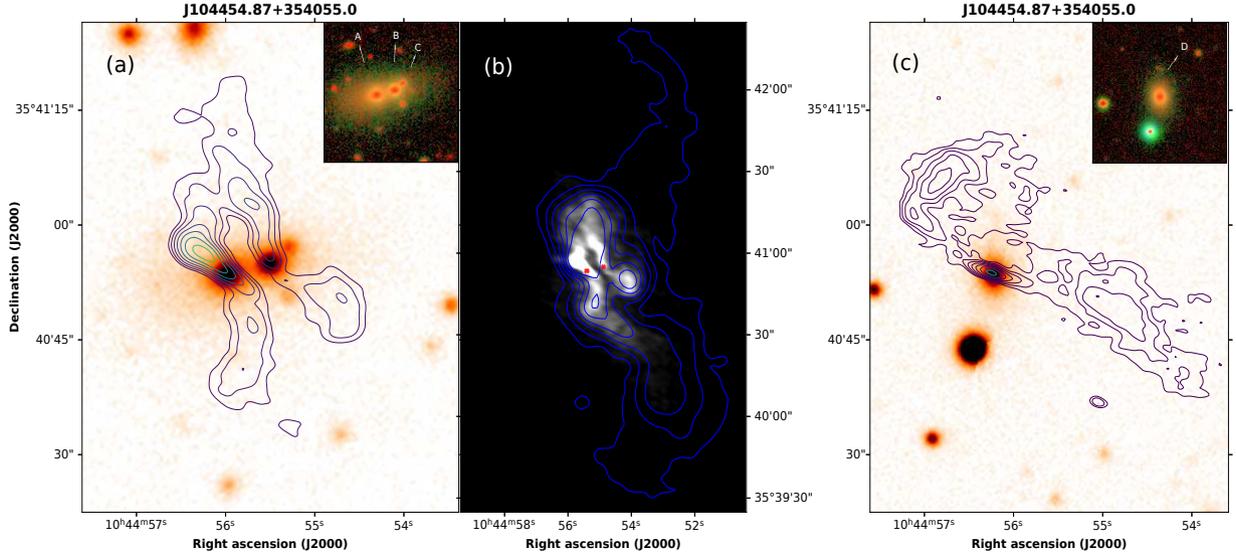}
 \caption{{\it Left Panel:}  The radio-optical overlay shows the $u$GMRT 1.4 GHz  radio contours overlaid on to the $r$ band image from the DECaLS \citep{Dey2019AJ....157..168D} of the dumb-bell galaxy pair `A' and `B'. The inset shows the $g$, $r$, $z$-band color-composite image from the DECaLS of the dumb-bell galaxy pair together with an apparently close radio-quiet  galaxy `C'.  {\it Middle panel:} shows the gray-scale uGMRT image of the pair of bi-polar jets ('Twin-Radio-Galaxy' TRG)  overlaid with the 144 MHz LoTSS-DR2 radio contours. The positions of the respective host galaxies are marked with red dots. {\it Right Panel:} Radio-optical overlay of the $u$GMRT 1.4 GHz radio contours of the `Wide-Angle-Tail' (WAT) radio galaxy `D' whose $g$, $r$, $z$-band color-composite DECaLS image is shown within the inset. Contours levels are set at square-root increment of (2, 4, 8, 16, 32, ...) $\times$ 14$\sigma$ for the TRG and $\times$ 5$\sigma$ for the WAT source, where $\sigma$ rms noise is 34 $\mu$Jy beam$^{-1}$ for the our  uGMRT maps. The LoTSS counter levels are set at  (1, 16, 81, 256, ...) $\times 3\sigma$, where $\sigma$ rms noise is 87 $\mu$Jy beam$^{-1}$.}
 \label{fig.trg_ugmrt}
\end{figure*}

\begin{figure}
    \includegraphics[width=8.0cm,height=5.5cm, trim={0.1cm 0 0.5cm 0},clip]{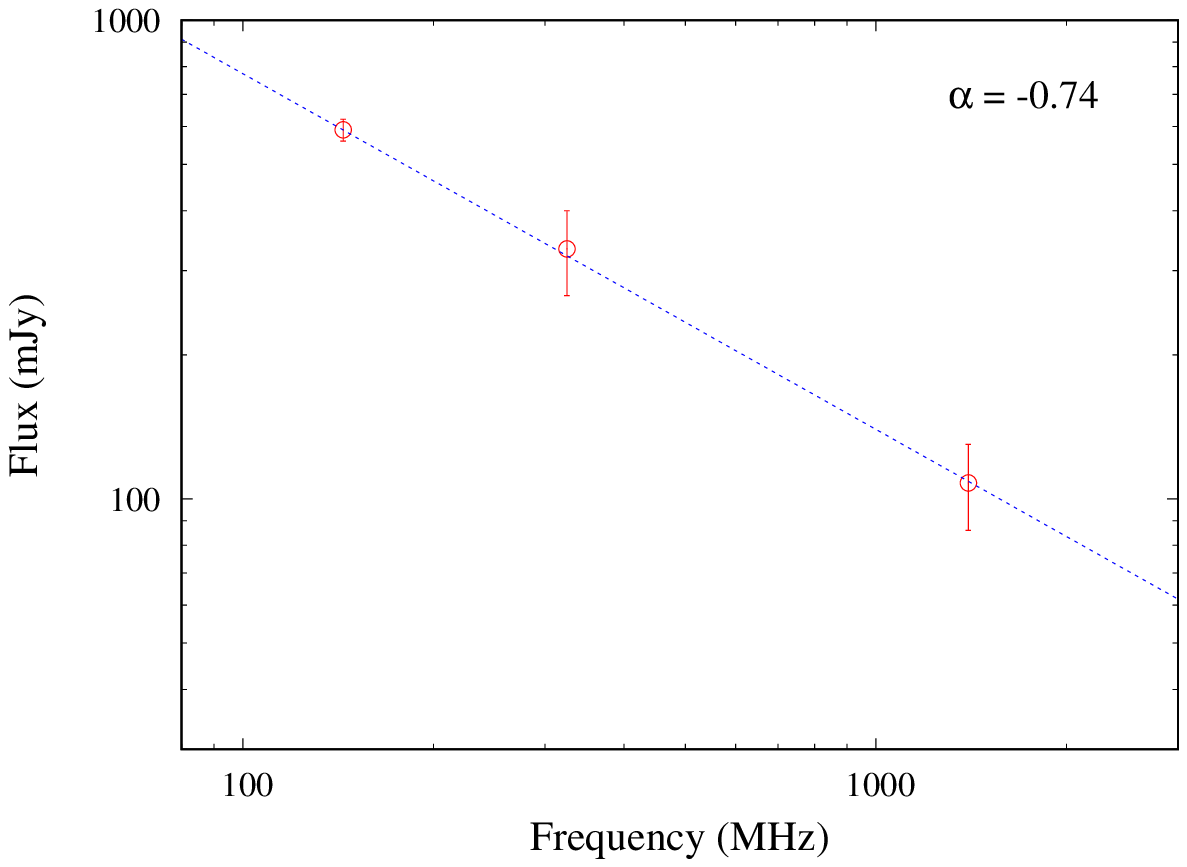}
 \caption{Integrated radio spectrum of the present TRG J1044+3540, using the flux densities from the 144 MHz LoTSS-DR2 \citep{Shimwell2022A&A...659A...1S},
325 MHz WENSS \citep{Rengelink1997A&AS..124..259R} and NVSS 1.4 GHz \citep{1998AJ....115.1693C}. 
The beamsize for the latter two maps is around 0.75 - 0.9 arcmin. Therefore, the LoTSS-DR2 map was smoothed to a beamwidth of 0.8 arcmin. The measured intergrated flux densities at the 3 frequencies are: $590 \pm 31$ mJy (144 MHz), $333 \pm 67$ mJy (325 MHz) and $108 \pm 22$ mJy (1.4 GHz).
 }
\label{fig:sindex}
\end{figure}

 In the uGMRT map, both members of the dumb-bell system, namely A $\&$ B (Table~\ref{tab:spec}) are seen to eject wiggling bi-polar radio jets along roughly the same direction  (Fig.~\ref{fig.trg_ugmrt}). Both northern jets are brighter than their southern counterparts and, after maintaining a steady mutual separation over the initial $\sim$ 30 kpc (projected), the two northern jets become diffuse but do not appear to merge (see, Section \ref{sect.discussion}). The (fainter) two jets on the southern side remain separated all along, despite strong wiggling. The observed projected extents of the bi-polar jets associated with A and B are $\sim$220 kpc, and $\sim$140 kpc, respectively  (Fig. \ref{fig.trg_ugmrt}).  The 144 MHz  radio maps from the LOFAR Two-meter Sky Survey Data Release 2 (LoTSS-DR2) \citep{Shimwell2022A&A...659A...1S}, has revealed more extended, bright radio emission connected to the southern jets, whereas 
a faint, long radio spur of size $\sim$ 60 arcsec is detected on the northern side (Fig. \ref{fig.trg_ugmrt}). Thus, the source has an overall size of at least 3.0 arcmin (i.e., 0.5 Mpc). However, the TRG morphology is not discernible in the LoTSS map, due to its relatively large beam size of $6 \times 6$ arcsec (see, Fig. ~\ref{fig.trg_ugmrt}). 
 Consistent with their moderate radio luminosities (Table~\ref{tab:spec}), both radio galaxies exhibit an edge-darkened (Fanaroff-Riley class I) morphology \citep{Fanaroff1974MNRAS.167P..31F, Ledlow1996AJ....112....9L}. 
 Fig.~\ref{fig:sindex} shows the integrated radio spectrum ( $S_{\nu} \propto \nu^{\alpha}$, where $\nu$ and $S_{\nu}$ are the frequency and flux density, respectively), of this dumb-bell system  (A + B), which is consistent with a straight line of spectral index $\alpha = - 0.74$. For the radio cores of the galaxies A and B, the VLASS maps \citep{Lacy2020PASP..132c5001L} give flux densities of $5.6 \pm 0.3 $ mJy, and $2.8\pm 0.3$ mJy at 3 GHz, respectively.  The galaxy C remained undetected in both VLASS and the present uGMRT observations, setting an upper limit of $1.4$ mJy at 1.4 GHz. The $u$GMRT map of the source D (Fig. \ref{fig.trg_ugmrt}) shows its radio jets emerging at PA $\approx$ 70 deg, although after $\sim$  20 arcsec ( $\simeq$ 56 kpc) the eastern jet undergoes a sharp bending towards the north-west, giving the source a `Wide-Angle-Tail' (WAT) morphology  \citep[see][]{Owen1976ApJ...205L...1O, Jones1979ApJ...234..818J}. From the VLASS, its core is found to have a flux density of $ 1.1\pm 0.2$ mJy at 3 GHz. Its integrated radio spectral index is $-0.75$, based on 
 the NVSS flux density of $23.4 \pm 1.3$ mJy at 1.4 GHz \citep{1998AJ....115.1693C} and  LoTSS-DR2 flux density of $127.7 \pm 12.2$ mJy at 144 MHz \citep{Shimwell2022A&A...659A...1S}.

\section{Discussion}
\label{sect.discussion}
Indubitably, a key message from the present study is that Twin-Radio-Galaxies (TRGs), each having a kiloparsec-scale bi-polar radio jet, are an extremely rare phenomenon.
The source J104454+354055 reported here becomes the third known TRG, the first one (3C 75) was reported 37 years ago and the second one (PKS 2149-158) 31 years ago.
All three TRGs are associated with dumb-bell systems residing in galaxy clusters/groups (Sect. \ref{sec_intro}). A prognosis of their extreme rarity came long ago from a VLA survey of 88 compact groups of galaxies \citep{Menon1985ApJ...296...60M}. These authors concluded that any bright radio emission arising from a galaxy group is usually associated with its brightest member galaxy, while the remaining group members are radio-quiet. They found a strong preference for the radio-loud elliptical galaxy in a group to be the first-ranked; even relatively luminous ellipticals holding second- or third- rank were found to be radio-quiet. A robust theoretical underpinning to this empirical result  still remains elusive.

Several parallels can be drawn between the present TRG and its two senior cousins, namely the famous and extensively studied 3C 75 and the TRG PKS 2149-158 (Sect. 1). As mentioned above, the galaxy pairs hosting all these 3 TRGs reside in a group/cluster environment, a circumstance clearly more conducive to the formation of dumb-bell systems. Secondly, all 3 TRGs exhibit FR I morphology, consistent with their moderate radio luminosities.
Their radio jets, imaged at 1.4 GHz with a similar resolution of $\sim$ 5 arcsec, too are similarly extended on 100 kpc scales. Also, the apparent separation between their twin host galaxies is around 10 kpc (between 8 and 18 kpc) and  
the line-of-sight velocity difference between the paired host galaxies are quite similar: $\sim 430\ \rm km\ s^{-1}$ (3C 75) \citep{Lauer..2014ApJ...797...82L}, $\sim 546\ \rm km\ s^{-1}$ (PKS 2149-158) \citep{Owen..1995AJ....109...14O, Smith..2004AJ....128.1558S}, and $\sim 300\ \rm km\ s^{-1}$ (present TRG) (see, Table~\ref{tab:spec}). 
However, beyond these basic similarities, salient differences  are noticeable between the 3 TRGs. Below, we enumerate some of these contrasts, as they might hold vital clues from the perspective of theoretical modelling and numerical simulations, especially once the other two TRGs have been observed in detail comparable to the proto-type 3C 75. 

(i) Whilst all 3 TRGs exhibit conspicuously wiggly jets, the two northern jets of 3C 75 are seen to merge into a single entity, fairly early on (a similar situation probably exists in the other TRG, PKS 2149-158, although even its best available (VLA) map lacks sufficient clarity, see \citep{Guidetti2008A&A...483..699G}. This contrasts with the present TRG in which both bi-polar jets retain their distinct identities much farther out, providing a greater space for their kinematical modelling.

(ii) In 3C 75,  situated  at the centre of the cluster Abell 400, all four jets undergo a sharp bending towards east/northeast \citep{Owen1985ApJ...294L..85O}.
This is attributed to dynamic pressure exerted by an intra-cluster  crosswind blowing at a speed of $\sim$ 1200 $\rm km\ s^{-1}$ \citep{Hudson2006A&A...453..433H}, due to the ongoing merger of the host cluster Abell 400 with a subcluster \citep[e.g.,][]{Beers1992ApJ...400..410B, Hudson2006A&A...453..433H}. This extraneous factor poses a complication to analytical modelling and numerical simulations of the basic phenomenon of twin bi-polar jets (e.g., \citealt{Molnar2017ApJ...835...57M, Musoke2020MNRAS.494.5207M}, see also, \citealt{Steinborn2016MNRAS.458.1013S}). A similar structural complexity appears to plague the twin bi-polar 
jets of the second TRG PKS 2149-158, which exhibit a conspicuous C-shaped bending \citep{Guidetti2008A&A...483..699G}, again indicating a strong dynamical pressure acting on the jets. In contrast, the present TRG is a much neater system, as the impact of the orbital motion of the two galaxies A and B on the morphologies of their bi-polar jets is not contaminated by ram-pressure of a cross-wind. This would simplify the dynamical modelling of the twin bi-polar jets, since there is no need to disentangle the effect of the orbital motion of the host galaxy pair from extraneous forces due to cross-wind impinging on the jets.
To appreciate this point, we may recall that the orbital motion of the paired host galaxies is expected to impose a mirror-symmetric wiggling on the bi-polar jet pair, as is indeed observed in the uGMRT image of the present system and which also lends further support to the assertion that the two bi-polar jets are physically associated (Fig. \ref{fig.trg_ugmrt}). In the other two TRGs, the wiggly pattern of jets is massively distorted due to the dynamical impact of the intracluster cross-wind, as mentioned above. The inference then is that the present TRG is a superior laboratory for probing the physics of two nearly parallel wiggly jets of relativistic plasma bouncing sideways against each other, but escaping disruption (Fig. \ref{fig.trg_ugmrt}). Also, all four jets of the present TRG are bright enough for radio polarimetry, which would enable mapping of their magnetic field and investigating its possible role in  preserving the identity of the relativistic plasma jets bouncing against each other.

\subsection{The WAT Neighbour}

As mentioned in Section~\ref{sect.res}, the $u$GMRT map also shows a WAT radio galaxy (D), located $2.72$ arcmin (= 456 kpc) to the west of the present TRG (Fig. ~\ref{fig.trg_ugmrt}c).
The twin-jets of the WAT initially advance along PA $\sim$ 70 deg, for $\sim$ 60 kpc, when the eastern jet undergoes a sharp bend towards northwest. 
Plausibly, the strong dynamic pressure required for the jet bending comes from motion of the host galaxy towards the dumb-bell system and its associated gaseous halo.
In order to assess such a possibility, we estimate the zone of gravitational influence of the dumb-bell, i.e., its virial radius, which is roughly the range out to which the gaseous halo of the dumb-bell system might extend.

 Using the 2MASS K-band apparent magnitudes of the dumb-bell constituents A and B, of $K = $ 14.48, and 14.89, respectively, we estimate their absolute magnitudes 
 ($M_K$) taking a distance modulus of 39.4 and K-corrections of $-0.27$ and $-0.28$ \citep[][]{Chilingarian2010MNRAS.405.1409C, Chilingarian2012MNRAS.419.1727C}.  
 This yields $M_K = -25.1$  and $-24.77$ for A and B, respectively.  These correspond to stellar masses of log $M_*$ $\sim$ 11.50$\ M_{\odot}$ and 11.35$\ M_{\odot}$ for A and B \citep[][equation 2]{Cappellari2013ApJ...778L...2C}, amounting to 
a total stellar mass of log $M_* \sim 11.73 M_\odot$ for the galaxy pair. 
Galaxy stellar-to-halo mass relation (SHMR) implies their dark halo virial masses to be log$M_{vir} \sim 14.1 M_{\odot}$  and  $\sim 13.8 M_{\odot}$, respectively 
(\citealt{Moster2010ApJ...710..903M},  equation 2; \citealt{Wechsler2018ARA&A..56..435W}). Such massive haloes are typically
assembled by a redshift of $z_a \sim 0.8$, and  have virial radii given by  \citet{Shull2014ApJ...784..142S}: 

\begin{equation}
 \frac{R_{\rm vir}}{ 206~ \rm kpc}  \approx 
\left(\frac{M_{\rm vir}}{10^{12}M_{\odot}}\right)^{1/3}  (1+z_a)^ {-1}
\label{eq1}
\end{equation}

For the total mass of the dumb-bell system (log$M_{vir} \sim 14.3 M_{\odot}$), its virial radius is then expected to extend to $\sim 670$ kpc, i.e., beyond the galaxy D identified with the WAT and is thus available for exerting the dynamic pressure required for the observed bending of its jet (Fig. ~\ref{fig.trg_ugmrt}c).

\section{Conclusions}

The $u$GMRT observations reported here have revealed of a pair of bi-polar extended radio jets emanating from a dumb-bell system containing two gravitationally bound massive elliptical galaxies within a common stellar envelope. This newly discovered `Twin-Radio-Galaxy’ (TRG), J104454+354055, is the third TRG found so far. The first 
(famous) example of this extremely rare class, the TRG 3C 75 was discovered 37 years ago, followed by the discovery of a second (much less known) TRG PKS 2149-158 three decades ago. All four jets of the present TRG are extended on 100-kiloparsec scale and exhibit an edge-darkened radio morphology (FR I), as well as a mirror-symmetric wiggly pattern which is consistent with both host galaxies performing orbital motion about a common gravitational centre (the recent LoTSS/DR2 map shows a total size of $\sim$0.5 Mpc).  Unlike the other two TRGs, there is no indication that the jets of the present TRG merge with each other early on, or become distorted due to a cross-wind. This goes to make this TRG a much neater exponent of cosmic tango, more amenable to modelling the jet dynamics in TRGs and for studying the impact of sideway collisions of relativistic plasma jets on their survivability
 \citep[e.g.,][]{Achterberg1988A&A...191..167A, Musoke2020MNRAS.494.5207M}.
A third bright radio galaxy in this group is found to be a Wide-Angle-Tail (WAT) apparently in motion directed towards the dumb-bell system. Radio spectral and polarimetric imaging of the present TRG, supplemented with X-ray observations yielding gas pressures within the group, and numerical simulations, would be very desirable for a deeper understanding of the evolution of closely interacting synchrotron jets. 

 \begin{table*}
 {\scriptsize
 \centering
 \begin{minipage}{180mm}
 \caption{Basic parameters of the 4 galaxies based on optical and radio $u$GMRT observations.} 
 \label{tab:spec}
 \begin{tabular}{@{} r c r r r r r c  c c c c c   @{}}
 \hline  \hline 
 Source     &    \multicolumn{1}{c}{R.A.}  &  \multicolumn{1}{c}{Dec.}    &  \multicolumn{1}{c}{redshift} \textcolor{blue}{$^\star$} &  \multicolumn{1}{c}{$\rm log \ M_{BH}$}   &  Type   &   \multicolumn{3}{c}{Flux density (mJy) }   &   \multicolumn{3}{c}{Length (arcsec) }   & Radio  Luminosity  \\
 	\cline{7	$\-$ -9}
		\cline{10	$\-$ -12}
          &    \multicolumn{1}{c}{(J2000)} & \multicolumn{1}{c}{(J2000)}    &   \multicolumn{1}{c}{$z$} & \multicolumn{1}{c}{$M_{\odot}$}           &   &    \multicolumn{1}{c}{Core \textcolor{blue}{$^\dagger$}}     &    \multicolumn{1}{c}{North-jet}            &   \multicolumn{1}{c}{South-jet}        &    \multicolumn{1}{c}{North-jet}        &     \multicolumn{1}{c}{South-jet}           &     \multicolumn{1}{c}{Total}  & log($L_R$) (erg/s)\\  
          \hline
A & $10^h 44^m 55.36^s$  & $+35^{\circ} 40^{'} 53.5^{''}$  & 0.16173 & 10.2$\pm0.4$ & LERG & $5.6 \pm 0.3$ & 21.8  & 40.8  & 29.7 & 51.3 & 78.9 & 42.43\\
B & $10^h 44^m 54.87^s$  & $+35^{\circ} 40^{'} 55.0^{''}$  & 0.16278 & 9.3$\pm$0.3 & LERG &  $2.8 \pm 0.3 $ & 36.1  & 23.5  & 31.7 & 20.0 & 51.3 & 42.36 \\
 C     & $10^h 44^m 54.67^s$  & $+35^{\circ} 40^{'} 57.1^{''}$  & $--$    & $--$          &  $--$&  $\le 1.4$ ($ 5\sigma$) & $--$  & $--$  & $--$ & $--$ & $--$ & $--$ \\
&&&&&&&&&&&&\\

\hline
         &                      &                            &         &               &      & \multicolumn{1}{c}{Core}     &    \multicolumn{1}{c}{West-jet}            &   \multicolumn{1}{c}{East-jet}        &    \multicolumn{1}{c}{West-jet}        &     \multicolumn{1}{c}{East-jet}           &     \multicolumn{1}{c}{Total}  & \\
D      & $10^h 44^m 41.85^s$  & $+35^{\circ} 40^{'} 16.7^{''}$  & 0.16007 &  9.4$\pm$0.2    & LERG &  $1.1 \pm 0.2$ & 19.5  & 20.4  & 46.6 & 26.1 & 73.8 &  41.87\\
\hline
 \hline               
 \end{tabular}  
 \\
 \textcolor{blue}{$^\dagger$} The core flux densities are estimated using the VLASS radio map at 2.5 GHz \citep{Lacy2020PASP..132c5001L}.  The other radio parameters are derived from 1.4 GHz uGMRT maps (present work).\\
\textcolor{blue}{$^\star$} The redshift of the galaxies B and D are taken from the SDSS archive \citep{Ahumada2020ApJS..249....3A}.
 \end{minipage}
 
 }
 \end{table*}

\section*{Acknowledgements}
GK acknowledges a Senior Scientist fellowship of the Indian National Science Academy (INSA). 
DP acknowledges the post-doctoral fellowship of the S.~N.~Bose National Centre for Basic Sciences, Kolkata, India, funded by the Department of Science and Technology (DST), India. XLY thanks the National Science Foundation of China (12103076), the Shanghai Sailing Program (21YF1455300) and the China Postdoctoral Science Foundation (2021M693267)for support, LCH was supported by the National Science Foundation of China (11721303, 11991052, 12011540375) and the China Manned Space Project (CMS-CSST-2021-A04). This research has made use of the ``K-corrections calculator'' service available at \href{http://kcor.sai.msu.ru/}{http://kcor.sai.msu.ru/}.
We thank the staff of the GMRT that made these observations possible. GMRT is run by the National Centre for Radio Astrophysics of the Tata Institute of Fundamental Research. We thank the staff of IAO, Hanle and CREST, Hosakote, that made these observations possible. The facilities at IAO and CREST are operated by the Indian Institute of Astrophysics, Bangalore.
%%%%%%%%%%%%%%%%%%%%%%%%%%%%%%%%%%%%%%%%%%%%%%%%%%
\section*{Data Availability}

The radio data used in this study will be publicly available at $u$GMRT data archive at \href{https://naps.ncra.tifr.res.in/goa/data/search}{https://naps.ncra.tifr.res.in/goa/data/search}

%%%%%%%%%%%%%%%%%%%% REFERENCES %%%%%%%%%%%%%%%%%%

% The best way to enter references is to use BibTeX:

\bibliographystyle{mnras}
% Alternatively you could enter them by hand, like this:
% This method is tedious and prone to error if you have lots of references
%\begin{thebibliography}{99}
%\bibitem[\protect\citeauthoryear{Author}{2012}]{Author2012}
%Author A.~N., 2013, Journal of Improbable Astronomy, 1, 1
%\bibitem[\protect\citeauthoryear{Others}{2013}]{Others2013}
%Others S., 2012, Journal of Interesting Stuff, 17, 198
%\end{thebibliography}

%%%%%%%%%%%%%%%%%%%%%%%%%%%%%%%%%%%%%%%%%%%%%%%%%%

%%%%%%%%%%%%%%%%% APPENDICES %%%%%%%%%%%%%%%%%%%%%

%%%%%%%%%%%%%%%%%%%%%%%%%%%%%%%%%%%%%%%%%%%%%%%%%%

% Don't change these lines
\bsp	% typesetting comment
\bibliography{references}

\label{lastpage}
\end{document}